\documentclass[aps,prl,twocolumn,groupedaddress,showpacs]{revtex4-1}

\usepackage{graphicx}

\begin{document}


\title{Formation of Ultracold Fermionic NaLi Feshbach Molecules}


\author{Myoung-Sun Heo$^1$, Tout T. Wang$^{1,2}$, Caleb A. Christensen$^1$, Timur M. Rvachov$^1$, Dylan A. Cotta$^{1,3}$ Jae-Hoon Choi$^1$, Ye-Ryoung Lee$^1$ and Wolfgang Ketterle$^1$}
\affiliation{$^1$MIT-Harvard Center for Ultracold Atoms, Research Laboratory of Electronics, Department of Physics, Massachusetts Institute of Technology, Cambridge, MA 02139, USA}
\affiliation{$^2$Department of Physics, Harvard University, Cambridge, MA 02138, USA}
\affiliation{$^3$D\'epartement de Physique, \'Ecole Normale Sup\'erieure de Cachan, 94235 Cachan, France}



\begin{abstract}
We describe the formation of fermionic NaLi Feshbach molecules from an ultracold mixture of bosonic $^{23}$Na and fermionic $^6$Li. Precise magnetic field sweeps across a narrow Feshbach resonance at $745$ G result in a molecule conversion fraction of $5\%$ for our experimental densities and temperatures, corresponding to a molecule number of $5\times 10^4$. The observed molecular decay lifetime is $1.3$ ms after removing free Li and Na atoms from the trap.
\end{abstract}

\pacs{67.85.-d, 34.50.-s, 05.30.Fk}

\date{\today}

\maketitle

%
%

The preparation and control of ultracold atoms has led to major advances in precision measurements and many-body physics. One current frontier is to extend this to diatomic molecules. Early experiments focused on homonuclear molecules, where highlights included the study of fermion pairs across the BEC-BCS crossover \cite{Varenna2008}. The preparation of heteronuclear molecules is more challenging because it requires a controlled reaction between two distinct atomic species. However, heteronuclear molecules can have a strong elecric dipole moment, which leads to a range of new scientific directions \cite{KremsBook}, including precision measurements, such as of the electron electric dipole moment \cite{HindsRev}, quantum computation mediated by dipolar coupling between molecular qubits \cite{DeMilleQuantComp} or in a hybrid system of molecules coupled to superconducting waveguides \cite{LukinMeso}, many-body physics with anisotropic long-range interactions \cite{BaranovRev,PfauDipRev}, and ultracold chemistry \cite{KremsColdChem}.

A number of experiments have explored molecule formation in ultracold atoms using photoassociation and Feshbach resonances \cite{KremsBook}. Due to the lower abundance of fermionic alkali isotopes, only one heteronuclear fermionic molecule $^{40}$K$^{87}$Rb has been produced at ultracold temperatures \cite{NiMol08}. Fermionic molecules are appealing due to Pauli suppression of $s$-wave collisions between identical fermions \cite{OspQuantContReact}, as well as prospects for preparing fermions with long-range interactions as a model system for electrons with Coulomb interactions \cite{BaranovRev}. In this paper, we report the formation of a new fermionic heteronuclear molecule $^{23}$Na$^6$Li.

NaLi has at least three unique features due to its constituents being the two smallest alkali atoms. First, its small reduced mass gives it a large rotational constant, which suppresses inelastic molecule-molecule collisions that occur via coupling between rotational levels \cite{DalgarnoSpinDepol}. Second, NaLi is reactive in its singlet $X^1\Sigma^+$ ground state, meaning that the reaction NaLi + NaLi $\to$ Na$_2$ + Li$_2$ is energetically allowed \cite{HutsonMolReact}, but with an unusually small predicted rate constant of $10^{-13}$ cm$^3$/s that is by far the lowest among all reactive heteronuclear alkali molecules \cite{JulUnivMolRates} and should allow lifetimes $>1$ s even without dipolar suppression \cite{MirandStereo}. This is related to NaLi having the smallest van der Waals C$_6$ coefficient of all heteronuclear alkali atom pairs \cite{FreconChemPhys}, which results in weak scattering by the long-range potential. Finally, this slow collision rate, together with weak spin-orbit coupling in diatomic molecules with small atomic numbers $Z$ of its constituents \cite{BernathBook}, may allow a long-lived triplet $a^3\Sigma^+$ ground-state in NaLi. This state has nonzero electric \emph{and} magnetic dipole moments, opening up the possibility of exploring physics ranging from magnetic \cite{KremsNJP} and electric field control \cite{KremsSpinRelax} of molecule-molecule collisions to realizing novel lattice spin Hamiltonians with coupling mediated by the electric dipole moment \cite{ZollerMol}. In addition to these three features, we note that NaLi has only a moderate dipole moment of $0.5$ Debye \cite{AymarDipMom} in its $X^1\Sigma^+$ ground state, comparable to KRb. Larger electric fields are required to align this dipole moment due to the large rotational constant  $B=0.4$ cm$^{-1}$ \cite{HesselNaLi}.

An earlier effort to produce NaLi Feshbach molecules was unsuccessful \cite{CalebThesis} due to incorrect assignments of interspecies Feshbach resonances between $^{23}$Na and $^6$Li \cite{StanNaLi, GacesaNaLi}, which predicted many resonances to be much stronger than in the recently revised assignments \cite{OberNaLi}. For molecule formation the relevant figure of merit of a Feshbach resonance is its energetic width $E_0$ \cite{Varenna2008}. The more familiar width $\Delta B$  characterizes the visibility of scattering length modification relative to the background scattering length $a_{bg}$ and is not directly relevant for molecule formation. $E_0$ and $\Delta B$ are related by $\Delta\mu \Delta B = \sqrt{E_0 \hbar^2/m a_{bg}^2}$, where $\Delta \mu$ is the differential magnetic moment between the atomic and molecular states and $m$ is the reduced mass of the two particle system. Previously, NaLi resonances observed below 1000 G were assigned to $l=0$ molecular bound states, where $l$ is the rotational angular momentum of the molecule. The strongest of these resonances had a predicted width $E_0/h=$ 6 kHz \cite{GacesaNaLi}. The revised interpretation \cite{OberNaLi} assigns these resonances to $l=2$ molecular bound states coupled to the $l=0$ open-channel via magnetic dipole-dipole interactions, with $E_0/h=$ 5 Hz for the strongest resonance, which is three orders of magnitude smaller.

This small $E_0$ results in a uniquely challenging situation for molecule formation. The Feshbach molecule has open-channel character only for magnetic fields $B$ around resonance $B_0$ such that $B-B_0 < E_0/\Delta \mu = 2$ $\mu$G \cite{ChinFesh}.  This is impossible to resolve experimentally, and precludes molecule formation via radio-frequency (RF) association \cite{ZirbelMolODT} or modulation of the magnetic field at a frequency resonant with the molecular binding energy \cite{ThompOscField}, both of which require a large open-channel contribution for good wave-function overlap between atomic and molecular states. This leaves magnetic field sweeps across resonance as the only feasible approach.
\begin{figure}\centering
\includegraphics[scale=0.6]{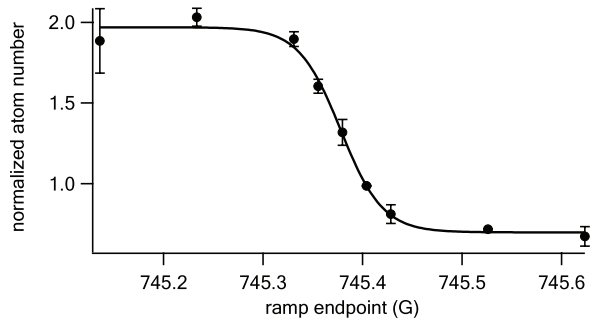}
\caption{Determination of Feshbach resonance location. Shown is the remaining atom number as a function of the endpoint of a downward magnetic field sweep across resonance. Results are fitted to a hyperbolic tangent function (solid line) with center 745.38 G and width 40 mG. The normalized atom number along the vertical axis is a sum of Na and Li atom numbers, each normalized to 1.}\label{fig:feshres}
\end{figure}

Such sweeps can be described as a Landau-Zener level-crossing of the molecular state and the unbound pair of atoms. In this simplified picture the molecule conversion fraction $f_\mathrm{mol} = 1 - e^{-2\pi \delta}$, with a Landau-Zener parameter $\delta$ given by \cite{Varenna2008,KohlerProdMol}
\begin{equation}\label{eq:LandauZener}
\delta = \frac{(\hbar \Omega)^2}{\hbar\Delta\mu\dot{B}}=\frac{4\pi\hbar n}{2m}\frac{\Delta B a_{bg}}{\dot{B}}=\frac{1}{3\pi}\frac{(\tilde{E}_F/E_0)^{3/2}}{\hbar\Delta\mu\dot{B}/E_0^2}
\end{equation}
where $\Omega=\frac{1}{\hbar}\sqrt{\tilde{E}_F^{3/2}E_0^{1/2}/3\pi}$ is the Rabi frequency for atom-molecule coupling, with an effective Fermi energy $\tilde{E}_F =(6 \pi^2 n)^{2/3}\hbar^2/m$ which expresses the atomic density $n$ in energy units for bosons as well as for fermions. $\Omega$ corresponds to the width of the level-crossing, and for efficient molecule formation $\delta > 1$, the magnetic field $B$ must be varied across a region of width $\hbar\Omega/\Delta\mu$ around resonance in a time $>1/\Omega$. In our experiment we use the strongest Feshbach resonance from the revised assignments \cite{OberNaLi}, located at $B_0 = 745$ G, with $\Delta B = 10$ mG, $a_{bg}=-70a_0$, $\Delta\mu = 2\mu_B$, $E_0/\Delta\mu$=2 $\mu$G as mentioned earlier, and $\hbar\Omega/\Delta\mu = 2$ mG for the experimental densities specified below. Working with such a narrow resonance at $745$ G requires careful stabilization of large magnetic fields. Molecule formation experiments with other comparably narrow resonances have been done, but at much lower fields, for example Cs$_2$ \cite{GrimmCs2Science} and $^6$Li$^{40}$K \cite{SpeigAllOpt}.

Experimentally we achieve $<$10 mG root-mean-square (rms) magnetic field noise. The power supply produces current fluctuations corresponding to $200$ mG of noise at $745$ G. Active feedback stabilization of the current reduces the noise to $<10$ mG rms after synchronizing experimental sequences with the 60 Hz line frequency.

We produce a near-degenerate mixture of $^{23}$Na atoms in the $^2S_{1/2}$ $\left| F, m_F \right> = \left| 2, 2 \right>$ state and $^6$Li atoms in the $^2S_{1/2}$  $\left| 3/2, 3/2 \right>$ state in a cloverleaf magnetic trap, using an apparatus similar to the one described in Ref. \cite{ZoranFifty}. The mixture is then transferred into a $5$ W single-beam optical dipole trap at $1064$  nm. In the optical trap we spin-flip both species with simultaneous Landau-Zener radio-frequency (RF) sweeps at a small bias field of $15$ G to the hyperfine ground states Na $\left| 1, 1 \right>$ and Li $\left| 1/2, 1/2 \right>$.

From low magnetic field we first jump far above the resonance to $752$ G, and after waiting $200$ ms the mixture is evaporatively cooled, leaving $1 \times 10^6$ each of Na and Li at a temperature of $1.2$ $\mu$K. Here, the weak magnetic field curvature along the optical dipole trap provides additional axial confinement, giving radial and axial trap frequencies $(\nu_r, \nu_z)$ of ($920$ Hz, $13$ Hz) for Na, and ($2.0$ kHz, $26$ Hz) for Li. The temperature of the mixture is near the onset of condensation for the bosonic Na, with $T/T_c = 1.2$, and $T/T_F = 0.3$, where for our trap parameters the condensation temperature for Na is $T_c = 1.0$ $\mu$K and the Fermi temperature for Li is $T_F = 4.1$ $\mu$K. The peak in-trap densities are $2.5 \times 10^{13}$ cm$^{-3}$ for Na and $1.4 \times 10^{13}$ cm$^{-3}$ for Li.
\begin{figure}\centering
\includegraphics[scale=0.4]{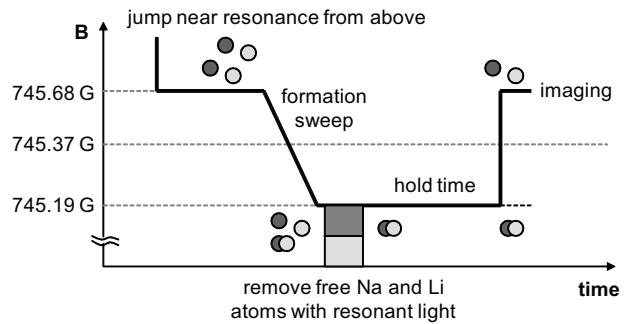}
\caption{Experimental sequence for molecule formation. After waiting for the magnetic field $B$ to stabilize at $745.68$ G, we sweep the field across resonance to $B=745.19$ G in $800$ $\mu$s, at a constant rate of $600$ mG/ms. At the end of the sweep, simultaneous pulses of resonant light remove Na and Li atoms from the trap within $100$ $\mu$s while leaving molecules unaffected. After a variable hold time for lifetime measurements, molecules are detected by jumping back above resonance and imaging dissociated atoms. The magnetic field can also be kept below resonance to confirm that bound molecules are invisible to the imaging light.}\label{fig:expseq}
\end{figure}

The precise location of the $745$ G resonance is determined by jumping down to $745.9$ G, waiting $15$ ms for the field to stabilize, and then sweeping the magnetic field down at a constant rate of $50$ mG/ms but with a variable end-point. The total atom number remaining after the sweep drops sharply as the end-point crosses the Feshbach resonance (Fig. \ref{fig:feshres}). This measurement determines the location of the Feshbach resonance to be $745.38\pm 0.04$ G, providing a precise target field for molecule formation sweeps. Note the $40$ mG width from Fig. \ref{fig:feshres} is not directly related to E$_0$, $\Omega$ or $\Delta B$.

A typical molecule formation sequence is shown in Fig. \ref{fig:expseq}. The start and end points of the magnetic field sweep are chosen to be as close to resonance as possible while still being outside the region of atom-molecule coupling defined by $\Omega$. This maximizes the formation efficiency, because molecules form and begin to decay immediately after crossing the coupling region around resonance.

At the end of the sweep, laser pulses resonant with the $589$ nm $^2S_{1/2}$ $\left| -1/2, 3/2 \right>$ $\rightarrow$ $^2P_{3/2}$ $\left| -3/2, 3/2 \right>$ transition in Na and the $671$ nm $^2S_{1/2}$ $\left| -1/2, 1 \right>$ $\rightarrow$ $^2P_{3/2}$ $\left| -3/2, 1 \right>$ transition in Li remove $>90\%$ of the remaining free atoms from the trap (states written in high-field basis $\left| m_J, m_I \right>$). For closed-channel Feshbach molecules, the excitation spectrum is sufficiently different from free atoms, so NaLi molecules are mostly unaffected if the laser pulses are not too long. After a variable hold time for molecule lifetime measurements, the field is switched above resonance, where bound molecules rapidly dissociate \cite{MukaiNaDissoc}. Imaging the free atoms gives us a measure of the molecule number, which for our optimized sweep parameters gives a formation fraction of $5\%$. This corresponds to a molecule number of $5\times 10^4$.
\begin{figure}\centering
\includegraphics[scale=0.45]{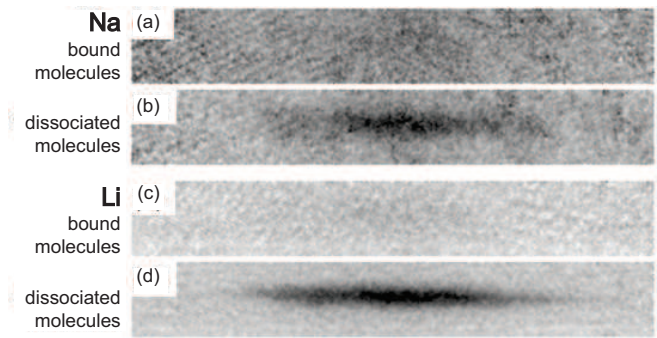}
\caption{Absorption imaging of molecular clouds using dissociated Na (a, b) and Li atoms (c, d). The lower images in each pair (b, d) are taken after switching above resonance and dissociating molecules into free atoms. The upper images (a, c) are reference images with the field below resonance, where free atoms have been removed from the trap and the molecules are invisible to the imaging light.}\label{fig:images}
\end{figure}

\begin{figure}\centering
\includegraphics[scale=0.4]{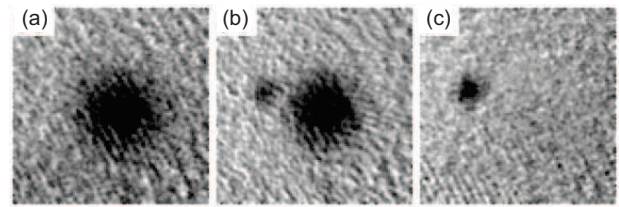}
\caption{Separation of atoms and molecules in a magnetic field gradient. Absorption images of (a) free Li atoms, (b) separated clouds of atoms and dissociated molecules, and (c) molecules with free atoms removed, taken $3$ ms after release from the crossed optical dipole trap in a $6$ G/cm gradient.}\label{fig:gradient}
\end{figure}

To confirm that the detected atomic signal is indeed from dissociated molecules, we check that imaging while keeping the magnetic field below resonance gives a negligible signal (Fig. \ref{fig:images}). Below resonance, bound molecules have a vanishing absorption cross section $\sigma$ and are invisible to the imaging light, while for free atoms $\sigma$ is nearly identical to above resonance because the Zeeman shift produced by magnetic field changes of $500$ mG is much smaller than the atomic linewidth. A second, independent confirmation of the molecular signal comes from releasing the trapped mixture in a magnetic field gradient, and demonstrating that the expanding atomic and molecular clouds separate from one another (Fig. \ref{fig:gradient}) because of their differential magnetic moment $\Delta\mu$. To observe this separation, we added additional optical confinement along the magnetic field gradient direction, using a second beam to form a crossed-beam optical dipole trap.

\begin{figure}\centering
\includegraphics[scale=0.6]{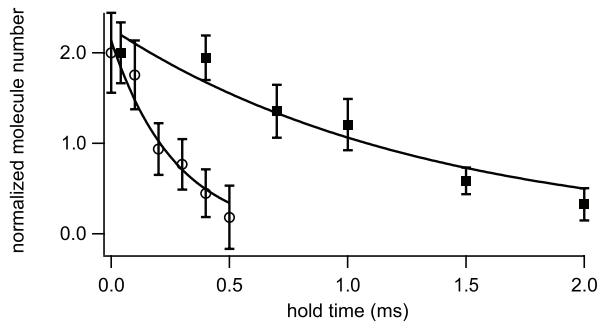}
\caption{Lifetime of trapped NaLi molecules after removing remaining free Na and Li atoms from the trap (solid squares). Fitting to an exponential gives a decay time constant of $1.3$ ms. For comparison, the lifetime without removing remaining free atoms from the trap is $270$ $\mu$s (open circles). The normalized molecule number along the vertical axis is a sum of Na and Li atom numbers, each normalized to 1 at zero hold time,  detected from dissociating bound NaLi molecules.}\label{fig:lifetime}
\end{figure}

We measure the lifetime of trapped NaLi molecules (Fig. \ref{fig:lifetime}) by varying the hold time before switching the magnetic field above resonance for dissociation and imaging. With $>90\%$ of both species of remaining atoms removed, the molecular lifetime is $1.3$ ms. This lifetime appears to be limited by collisions with other molecules or leftover atoms rather than by photon scattering, since it does not increase significantly with reduced intensity of the trapping laser. It can be enhanced by suppressing molecule-molecule collisions in a 3D optical lattice \cite{ChotiaLattice} and fully removing residual free atoms from the trap. The molecular lifetime drops to $270$ $\mu$s when keeping free Na and Li atoms trapped with the molecules. Finally, if only Li atoms are removed before the hold time and not Na, the lifetime is $550$ $\mu$s. The presence of free atoms increases the molecular decay rate because of inelastic collisions with molecules. Our lifetime measurements show that Na and Li each give comparable contributions to this increased decay, which is consistent with the constituent atoms in the closed-channel NaLi molecule being distinguishable from free Na and Li atoms, meaning that quantum statistics does not play a role in collisions. In contrast, experiments with open-channel KRb molecules \cite{ZirbelFeshStab} showed a sharp dependence of lifetime on the quantum statistics of the atomic collisional partner. The lifetimes above can also be reported as two-body loss-rate constants $\beta_\mathrm{NaLi+Na}\sim 1\times10^{-10}$ cm$^3$/s and $\beta_\mathrm{NaLi+Li}\sim 4\times10^{-10}$ cm$^3$/s.

The Landau-Zener parameter in our experiment is estimated to be $\delta=0.13$ by choosing $n$ in Eq. \ref{eq:LandauZener} to be the larger of $n_\mathrm{Na}$ and $n_\mathrm{Li}$. This corresponds to $f_\mathrm{mol}=56\%$. The discrepancy with the observed molecule fraction of $5\%$ has several explanations. The simplified Landau-Zener picture above assumes full phase space overlap between the atoms involved in molecule formation, which is only true in a Bose-Einstein condensate at $T=0$ \cite{HodbyPhaseSp,WillMolTheory}. In a Bose-Fermi mixture like NaLi, the phase space overlap is lower, reaching a maximum around the boson condensation temperature $T_c$ because at lower $T$ the Na condensate begins to have less spatial overlap with the fermionic Li \cite{ZirbelMolODT}. The phase-space overlap between Na and Li is further reduced because of their different trapping potentials \cite{DeuretzMolOptLatt}. Finally, the optimized sweep time of $800$ $\mu$s is much longer than the $270$ $\mu$s lifetime of NaLi molecules when trapped with free atoms, meaning that $80\%$ of molecules formed are lost, assuming that molecule formation occurs at the midpoint of the sweep.

In summary, we have succeeded in forming fermionic $^{23}$Na$^6$Li molecules around a narrow, closed-channel dominated Feshbach resonance at $745$ G. Optimized magnetic field sweeps across resonance result in a molecule conversion fraction of $5\%$ for our experimental densities and temperatures, corresponding to a molecule number of $5\times 10^4$. The observed molecular decay lifetime of $1.3$ ms is enough to begin exploring routes for reaching the ro-vibrational ground state of the molecule by stimulated Raman transitions \cite{NiMol08}, and to explore the unique features of this smallest heteronuclear alkali molecule.

\begin{acknowledgments}
This work was supported an AFOSR MURI on Ultracold Molecules, by the NSF and the ONR, and by ARO grant no. W911NF-07-1-0493 with funds from the DARPA Optical Lattice Emulator program. T.T.W. acknowledges additional support from NSERC. We are grateful to Jook Walraven, Tobias Tiecke, Roman Krems, John Doyle, and Dave Pritchard for valuable discussions, and also Colin Kennedy and Gregory Lau for experimental assistance.
\end{acknowledgments}


\begin{thebibliography}{38}%
\makeatletter
\providecommand \@ifxundefined [1]{%
 \@ifx{#1\undefined}
}%
\providecommand \@ifnum [1]{%
 \ifnum #1\expandafter \@firstoftwo
 \else \expandafter \@secondoftwo
 \fi
}%
\providecommand \@ifx [1]{%
 \ifx #1\expandafter \@firstoftwo
 \else \expandafter \@secondoftwo
 \fi
}%
\providecommand \natexlab [1]{#1}%
\providecommand \enquote  [1]{``#1''}%
\providecommand \bibnamefont  [1]{#1}%
\providecommand \bibfnamefont [1]{#1}%
\providecommand \citenamefont [1]{#1}%
\providecommand \href@noop [0]{\@secondoftwo}%
\providecommand \href [0]{\begingroup \@sanitize@url \@href}%
\providecommand \@href[1]{\@@startlink{#1}\@@href}%
\providecommand \@@href[1]{\endgroup#1\@@endlink}%
\providecommand \@sanitize@url [0]{\catcode `\\12\catcode `\$12\catcode
  `\&12\catcode `\#12\catcode `\^12\catcode `\_12\catcode `\%12\relax}%
\providecommand \@@startlink[1]{}%
\providecommand \@@endlink[0]{}%
\providecommand \url  [0]{\begingroup\@sanitize@url \@url }%
\providecommand \@url [1]{\endgroup\@href {#1}{\urlprefix }}%
\providecommand \urlprefix  [0]{URL }%
\providecommand \Eprint [0]{\href }%
\providecommand \doibase [0]{http://dx.doi.org/}%
\providecommand \selectlanguage [0]{\@gobble}%
\providecommand \bibinfo  [0]{\@secondoftwo}%
\providecommand \bibfield  [0]{\@secondoftwo}%
\providecommand \translation [1]{[#1]}%
\providecommand \BibitemOpen [0]{}%
\providecommand \bibitemStop [0]{}%
\providecommand \bibitemNoStop [0]{.\EOS\space}%
\providecommand \EOS [0]{\spacefactor3000\relax}%
\providecommand \BibitemShut  [1]{\csname bibitem#1\endcsname}%
\let\auto@bib@innerbib\@empty
\bibitem [{\citenamefont {Ketterle}\ and\ \citenamefont
  {Zwierlein}(2008)}]{Varenna2008}%
  \BibitemOpen
  \bibfield  {author} {\bibinfo {author} {\bibfnamefont {W.}~\bibnamefont
  {Ketterle}}\ and\ \bibinfo {author} {\bibfnamefont {M.~W.}\ \bibnamefont
  {Zwierlein}},\ }\href@noop {} {\emph {\bibinfo {title} {Ultracold Fermi
  Gases, Proceedings of the International School of Physics Enrico Fermi,
  Course CLXIV}}},\ edited by\ \bibinfo {editor} {\bibfnamefont
  {M.}~\bibnamefont {Inguscio}}, \bibinfo {editor} {\bibfnamefont
  {W.}~\bibnamefont {Ketterle}}, \ and\ \bibinfo {editor} {\bibfnamefont
  {C.}~\bibnamefont {Salomon}}\ (\bibinfo  {publisher} {IOS Press, Amsterdam},\
  \bibinfo {year} {2008})\BibitemShut {NoStop}%
\bibitem [{\citenamefont {Krems}\ \emph {et~al.}(2009)\citenamefont {Krems},
  \citenamefont {Stwalley},\ and\ \citenamefont {Friedrich}}]{KremsBook}%
  \BibitemOpen
  \bibinfo {editor} {\bibfnamefont {R.~V.}\ \bibnamefont {Krems}}, \bibinfo
  {editor} {\bibfnamefont {W.~C.}\ \bibnamefont {Stwalley}}, \ and\ \bibinfo
  {editor} {\bibfnamefont {B.}~\bibnamefont {Friedrich}},\ eds.,\ \href@noop {}
  {\emph {\bibinfo {title} {Cold Molecules: Theory, Experiment,
  Applications}}}\ (\bibinfo  {publisher} {CRC Press, Boca Raton, FL},\
  \bibinfo {year} {2009})\BibitemShut {NoStop}%
\bibitem [{\citenamefont {Hinds}(1997)}]{HindsRev}%
  \BibitemOpen
  \bibfield  {author} {\bibinfo {author} {\bibfnamefont {E.~A.}\ \bibnamefont
  {Hinds}},\ }\href {\doibase doi:10.1088/0031-8949/1997/T70/005} {\bibfield
  {journal} {\bibinfo  {journal} {Phys. Scripta}\ }\textbf {\bibinfo {volume}
  {T70}},\ \bibinfo {pages} {34} (\bibinfo {year} {1997})}\BibitemShut
  {NoStop}%
\bibitem [{\citenamefont {DeMille}(2002)}]{DeMilleQuantComp}%
  \BibitemOpen
  \bibfield  {author} {\bibinfo {author} {\bibfnamefont {D.}~\bibnamefont
  {DeMille}},\ }\href {\doibase 10.1103/PhysRevLett.88.067901} {\bibfield
  {journal} {\bibinfo  {journal} {Phys. Rev. Lett.}\ }\textbf {\bibinfo
  {volume} {88}},\ \bibinfo {pages} {067901} (\bibinfo {year}
  {2002})}\BibitemShut {NoStop}%
\bibitem [{\citenamefont {Andr\'e}\ \emph {et~al.}(2006)\citenamefont
  {Andr\'e}, \citenamefont {DeMille}, \citenamefont {Doyle}, \citenamefont
  {Lukin}, \citenamefont {Maxwell}, \citenamefont {Rabl}, \citenamefont
  {Schoelkopf},\ and\ \citenamefont {Zoller}}]{LukinMeso}%
  \BibitemOpen
  \bibfield  {author} {\bibinfo {author} {\bibfnamefont {A.}~\bibnamefont
  {Andr\'e}}, \bibinfo {author} {\bibfnamefont {D.}~\bibnamefont {DeMille}},
  \bibinfo {author} {\bibfnamefont {J.~M.}\ \bibnamefont {Doyle}}, \bibinfo
  {author} {\bibfnamefont {M.~D.}\ \bibnamefont {Lukin}}, \bibinfo {author}
  {\bibfnamefont {S.~E.}\ \bibnamefont {Maxwell}}, \bibinfo {author}
  {\bibfnamefont {P.}~\bibnamefont {Rabl}}, \bibinfo {author} {\bibfnamefont
  {R.~J.}\ \bibnamefont {Schoelkopf}}, \ and\ \bibinfo {author} {\bibfnamefont
  {P.}~\bibnamefont {Zoller}},\ }\href {\doibase 10.1038/nphys386} {\bibfield
  {journal} {\bibinfo  {journal} {Nature Phys.}\ }\textbf {\bibinfo {volume}
  {2}},\ \bibinfo {pages} {636} (\bibinfo {year} {2006})}\BibitemShut {NoStop}%
\bibitem [{\citenamefont {Baranov}(2008)}]{BaranovRev}%
  \BibitemOpen
  \bibfield  {author} {\bibinfo {author} {\bibfnamefont {M.~A.}\ \bibnamefont
  {Baranov}},\ }\href {\doibase 10.1016/j.physrep.2008.04.007} {\bibfield
  {journal} {\bibinfo  {journal} {Phys. Rep.}\ }\textbf {\bibinfo {volume}
  {464}},\ \bibinfo {pages} {71} (\bibinfo {year} {2008})}\BibitemShut
  {NoStop}%
\bibitem [{\citenamefont {Lahaye}\ \emph {et~al.}(2009)\citenamefont {Lahaye},
  \citenamefont {Menotti}, \citenamefont {Santos}, \citenamefont {Lewenstein},\
  and\ \citenamefont {Pfau}}]{PfauDipRev}%
  \BibitemOpen
  \bibfield  {author} {\bibinfo {author} {\bibfnamefont {T.}~\bibnamefont
  {Lahaye}}, \bibinfo {author} {\bibfnamefont {C.}~\bibnamefont {Menotti}},
  \bibinfo {author} {\bibfnamefont {L.}~\bibnamefont {Santos}}, \bibinfo
  {author} {\bibfnamefont {M.}~\bibnamefont {Lewenstein}}, \ and\ \bibinfo
  {author} {\bibfnamefont {T.}~\bibnamefont {Pfau}},\ }\href {\doibase
  doi:10.1088/0034-4885/72/12/126401} {\bibfield  {journal} {\bibinfo
  {journal} {Rep. Prog. Phys.}\ }\textbf {\bibinfo {volume} {72}},\ \bibinfo
  {pages} {126401} (\bibinfo {year} {2009})}\BibitemShut {NoStop}%
\bibitem [{\citenamefont {Krems}(2008)}]{KremsColdChem}%
  \BibitemOpen
  \bibfield  {author} {\bibinfo {author} {\bibfnamefont {R.~V.}\ \bibnamefont
  {Krems}},\ }\href {\doibase 10.1039/B802322K} {\bibfield  {journal} {\bibinfo
   {journal} {Phys. Chem. Chem. Phys.}\ }\textbf {\bibinfo {volume} {10}},\
  \bibinfo {pages} {4079} (\bibinfo {year} {2008})}\BibitemShut {NoStop}%
\bibitem [{\citenamefont {Ni}\ \emph {et~al.}(2008)\citenamefont {Ni},
  \citenamefont {Ospelkaus}, \citenamefont {de~Miranda}, \citenamefont {Pe'er},
  \citenamefont {Neyenhuis}, \citenamefont {Zirbel}, \citenamefont
  {Kotochigova}, \citenamefont {Julienne}, \citenamefont {Jin},\ and\
  \citenamefont {Ye}}]{NiMol08}%
  \BibitemOpen
  \bibfield  {author} {\bibinfo {author} {\bibfnamefont {K.-K.}\ \bibnamefont
  {Ni}}, \bibinfo {author} {\bibfnamefont {S.}~\bibnamefont {Ospelkaus}},
  \bibinfo {author} {\bibfnamefont {M.~H.~G.}\ \bibnamefont {de~Miranda}},
  \bibinfo {author} {\bibfnamefont {A.}~\bibnamefont {Pe'er}}, \bibinfo
  {author} {\bibfnamefont {B.}~\bibnamefont {Neyenhuis}}, \bibinfo {author}
  {\bibfnamefont {J.~J.}\ \bibnamefont {Zirbel}}, \bibinfo {author}
  {\bibfnamefont {S.}~\bibnamefont {Kotochigova}}, \bibinfo {author}
  {\bibfnamefont {P.~S.}\ \bibnamefont {Julienne}}, \bibinfo {author}
  {\bibfnamefont {D.~S.}\ \bibnamefont {Jin}}, \ and\ \bibinfo {author}
  {\bibfnamefont {J.}~\bibnamefont {Ye}},\ }\href {\doibase
  10.1126/science.1163861} {\bibfield  {journal} {\bibinfo  {journal}
  {Science}\ }\textbf {\bibinfo {volume} {322}},\ \bibinfo {pages} {231}
  (\bibinfo {year} {2008})}\BibitemShut {NoStop}%
\bibitem [{\citenamefont {Ospelkaus}\ \emph {et~al.}(2010)\citenamefont
  {Ospelkaus}, \citenamefont {Ni}, \citenamefont {Wang}, \citenamefont
  {de~Miranda}, \citenamefont {Neyenhuis}, \citenamefont {Qu\'{e}m\'{e}ner},
  \citenamefont {Julienne}, \citenamefont {Bohn}, \citenamefont {Jin},\ and\
  \citenamefont {Ye}}]{OspQuantContReact}%
  \BibitemOpen
  \bibfield  {author} {\bibinfo {author} {\bibfnamefont {S.}~\bibnamefont
  {Ospelkaus}}, \bibinfo {author} {\bibfnamefont {K.-K.}\ \bibnamefont {Ni}},
  \bibinfo {author} {\bibfnamefont {D.}~\bibnamefont {Wang}}, \bibinfo {author}
  {\bibfnamefont {M.~H.~G.}\ \bibnamefont {de~Miranda}}, \bibinfo {author}
  {\bibfnamefont {B.}~\bibnamefont {Neyenhuis}}, \bibinfo {author}
  {\bibfnamefont {G.}~\bibnamefont {Qu\'{e}m\'{e}ner}}, \bibinfo {author}
  {\bibfnamefont {P.~S.}\ \bibnamefont {Julienne}}, \bibinfo {author}
  {\bibfnamefont {J.~L.}\ \bibnamefont {Bohn}}, \bibinfo {author}
  {\bibfnamefont {D.~S.}\ \bibnamefont {Jin}}, \ and\ \bibinfo {author}
  {\bibfnamefont {J.}~\bibnamefont {Ye}},\ }\href {\doibase
  10.1126/science.1184121} {\bibfield  {journal} {\bibinfo  {journal}
  {Science}\ }\textbf {\bibinfo {volume} {327}},\ \bibinfo {pages} {853}
  (\bibinfo {year} {2010})}\BibitemShut {NoStop}%
\bibitem [{\citenamefont {Krems}\ and\ \citenamefont
  {Dalgarno}(2004)}]{DalgarnoSpinDepol}%
  \BibitemOpen
  \bibfield  {author} {\bibinfo {author} {\bibfnamefont {R.~V.}\ \bibnamefont
  {Krems}}\ and\ \bibinfo {author} {\bibfnamefont {A.}~\bibnamefont
  {Dalgarno}},\ }\href {\doibase 10.1063/1.1636691} {\bibfield  {journal}
  {\bibinfo  {journal} {J. Chem. Phys.}\ }\textbf {\bibinfo {volume} {120}},\
  \bibinfo {pages} {2296} (\bibinfo {year} {2004})}\BibitemShut {NoStop}%
\bibitem [{\citenamefont {\ifmmode~\dot{Z}\else \.{Z}\fi{}uchowski}\ and\
  \citenamefont {Hutson}(2010)}]{HutsonMolReact}%
  \BibitemOpen
  \bibfield  {author} {\bibinfo {author} {\bibfnamefont {P.~S.}\ \bibnamefont
  {\ifmmode~\dot{Z}\else \.{Z}\fi{}uchowski}}\ and\ \bibinfo {author}
  {\bibfnamefont {J.~M.}\ \bibnamefont {Hutson}},\ }\href {\doibase
  10.1103/PhysRevA.81.060703} {\bibfield  {journal} {\bibinfo  {journal} {Phys.
  Rev. A}\ }\textbf {\bibinfo {volume} {81}},\ \bibinfo {pages} {060703}
  (\bibinfo {year} {2010})}\BibitemShut {NoStop}%
\bibitem [{\citenamefont {Julienne}\ \emph {et~al.}(2011)\citenamefont
  {Julienne}, \citenamefont {Hanna},\ and\ \citenamefont
  {Idziaszek}}]{JulUnivMolRates}%
  \BibitemOpen
  \bibfield  {author} {\bibinfo {author} {\bibfnamefont {P.~S.}\ \bibnamefont
  {Julienne}}, \bibinfo {author} {\bibfnamefont {T.~M.}\ \bibnamefont {Hanna}},
  \ and\ \bibinfo {author} {\bibfnamefont {Z.}~\bibnamefont {Idziaszek}},\
  }\href {\doibase 10.1039/C1CP21270B} {\bibfield  {journal} {\bibinfo
  {journal} {Phys. Chem. Chem. Phys.}\ }\textbf {\bibinfo {volume} {13}},\
  \bibinfo {pages} {19114} (\bibinfo {year} {2011})}\BibitemShut {NoStop}%
\bibitem [{\citenamefont {de~Miranda}\ \emph {et~al.}(2011)\citenamefont
  {de~Miranda}, \citenamefont {Chotia}, \citenamefont {Neyenhuis},
  \citenamefont {Wang}, \citenamefont {Qu\'{e}m\'{e}ner}, \citenamefont
  {Ospelkaus}, \citenamefont {Bohn}, \citenamefont {Ye},\ and\ \citenamefont
  {Jin}}]{MirandStereo}%
  \BibitemOpen
  \bibfield  {author} {\bibinfo {author} {\bibfnamefont {M.~H.~G.}\
  \bibnamefont {de~Miranda}}, \bibinfo {author} {\bibfnamefont
  {A.}~\bibnamefont {Chotia}}, \bibinfo {author} {\bibfnamefont
  {B.}~\bibnamefont {Neyenhuis}}, \bibinfo {author} {\bibfnamefont
  {D.}~\bibnamefont {Wang}}, \bibinfo {author} {\bibfnamefont {G.}~\bibnamefont
  {Qu\'{e}m\'{e}ner}}, \bibinfo {author} {\bibfnamefont {S.}~\bibnamefont
  {Ospelkaus}}, \bibinfo {author} {\bibfnamefont {J.~L.}\ \bibnamefont {Bohn}},
  \bibinfo {author} {\bibfnamefont {J.}~\bibnamefont {Ye}}, \ and\ \bibinfo
  {author} {\bibfnamefont {D.~S.}\ \bibnamefont {Jin}},\ }\href {\doibase
  10.1038/nphys1939} {\bibfield  {journal} {\bibinfo  {journal} {Nature Phys.}\
  }\textbf {\bibinfo {volume} {7}},\ \bibinfo {pages} {502} (\bibinfo {year}
  {2011})}\BibitemShut {NoStop}%
\bibitem [{\citenamefont {Bussery}\ \emph {et~al.}(1987)\citenamefont
  {Bussery}, \citenamefont {Achkar},\ and\ \citenamefont
  {Aubert-Fr\'econ}}]{FreconChemPhys}%
  \BibitemOpen
  \bibfield  {author} {\bibinfo {author} {\bibfnamefont {B.}~\bibnamefont
  {Bussery}}, \bibinfo {author} {\bibfnamefont {Y.}~\bibnamefont {Achkar}}, \
  and\ \bibinfo {author} {\bibfnamefont {M.}~\bibnamefont {Aubert-Fr\'econ}},\
  }\href {\doibase 10.1016/0301-0104(87)80202-1} {\bibfield  {journal}
  {\bibinfo  {journal} {Chem. Phys.}\ }\textbf {\bibinfo {volume} {116}},\
  \bibinfo {pages} {319} (\bibinfo {year} {1987})}\BibitemShut {NoStop}%
\bibitem [{\citenamefont {Bernath}(2005)}]{BernathBook}%
  \BibitemOpen
  \bibfield  {author} {\bibinfo {author} {\bibfnamefont {P.~F.}\ \bibnamefont
  {Bernath}},\ }\href@noop {} {\emph {\bibinfo {title} {Spectra of Atoms and
  Molecules, 2nd Ed.}}}\ (\bibinfo  {publisher} {Oxford Univ. Press},\ \bibinfo
  {year} {2005})\BibitemShut {NoStop}%
\bibitem [{\citenamefont {Tscherbul}\ \emph {et~al.}(2009)\citenamefont
  {Tscherbul}, \citenamefont {Suleimanov}, \citenamefont {Aquilanti},\ and\
  \citenamefont {Krems}}]{KremsNJP}%
  \BibitemOpen
  \bibfield  {author} {\bibinfo {author} {\bibfnamefont {T.~V.}\ \bibnamefont
  {Tscherbul}}, \bibinfo {author} {\bibfnamefont {Y.~V.}\ \bibnamefont
  {Suleimanov}}, \bibinfo {author} {\bibfnamefont {V.}~\bibnamefont
  {Aquilanti}}, \ and\ \bibinfo {author} {\bibfnamefont {R.~V.}\ \bibnamefont
  {Krems}},\ }\href {\doibase 10.1088/1367-2630/11/5/055021} {\bibfield
  {journal} {\bibinfo  {journal} {New J. Phys.}\ }\textbf {\bibinfo {volume}
  {11}},\ \bibinfo {pages} {055021} (\bibinfo {year} {2009})}\BibitemShut
  {NoStop}%
\bibitem [{\citenamefont {Tscherbul}\ and\ \citenamefont
  {Krems}(2006)}]{KremsSpinRelax}%
  \BibitemOpen
  \bibfield  {author} {\bibinfo {author} {\bibfnamefont {T.~V.}\ \bibnamefont
  {Tscherbul}}\ and\ \bibinfo {author} {\bibfnamefont {R.~V.}\ \bibnamefont
  {Krems}},\ }\href {\doibase 10.1103/PhysRevLett.97.083201} {\bibfield
  {journal} {\bibinfo  {journal} {Phys. Rev. Lett.}\ }\textbf {\bibinfo
  {volume} {97}},\ \bibinfo {pages} {083201} (\bibinfo {year}
  {2006})}\BibitemShut {NoStop}%
\bibitem [{\citenamefont {Micheli}\ \emph {et~al.}(2006)\citenamefont
  {Micheli}, \citenamefont {Brennen},\ and\ \citenamefont
  {Zoller}}]{ZollerMol}%
  \BibitemOpen
  \bibfield  {author} {\bibinfo {author} {\bibfnamefont {A.}~\bibnamefont
  {Micheli}}, \bibinfo {author} {\bibfnamefont {G.~K.}\ \bibnamefont
  {Brennen}}, \ and\ \bibinfo {author} {\bibfnamefont {P.}~\bibnamefont
  {Zoller}},\ }\href {\doibase 10.1038/nphys287} {\bibfield  {journal}
  {\bibinfo  {journal} {Nature Phys.}\ }\textbf {\bibinfo {volume} {2}},\
  \bibinfo {pages} {341} (\bibinfo {year} {2006})}\BibitemShut {NoStop}%
\bibitem [{\citenamefont {Aymar}\ and\ \citenamefont
  {Dulieu}(2005)}]{AymarDipMom}%
  \BibitemOpen
  \bibfield  {author} {\bibinfo {author} {\bibfnamefont {M.}~\bibnamefont
  {Aymar}}\ and\ \bibinfo {author} {\bibfnamefont {O.}~\bibnamefont {Dulieu}},\
  }\href {\doibase 10.1063/1.1903944} {\bibfield  {journal} {\bibinfo
  {journal} {J. Chem. Phys.}\ }\textbf {\bibinfo {volume} {122}},\ \bibinfo
  {pages} {204302} (\bibinfo {year} {2005})}\BibitemShut {NoStop}%
\bibitem [{\citenamefont {Hessel}(1971)}]{HesselNaLi}%
  \BibitemOpen
  \bibfield  {author} {\bibinfo {author} {\bibfnamefont {M.~M.}\ \bibnamefont
  {Hessel}},\ }\href {\doibase 10.1103/PhysRevLett.26.215} {\bibfield
  {journal} {\bibinfo  {journal} {Phys. Rev. Lett.}\ }\textbf {\bibinfo
  {volume} {26}},\ \bibinfo {pages} {215} (\bibinfo {year} {1971})}\BibitemShut
  {NoStop}%
\bibitem [{\citenamefont {Christensen}(2011)}]{CalebThesis}%
  \BibitemOpen
  \bibfield  {author} {\bibinfo {author} {\bibfnamefont {C.~A.}\ \bibnamefont
  {Christensen}},\ }\emph {\bibinfo {title} {Ultracold Molecules from Ultracold
  Atoms: Interactions in Sodium and Lithium Gas}},\ \href@noop {} {Ph.D.
  thesis},\ \bibinfo  {school} {Massachusetts Institute of Technology}
  (\bibinfo {year} {2011})\BibitemShut {NoStop}%
\bibitem [{\citenamefont {Stan}\ \emph {et~al.}(2004)\citenamefont {Stan},
  \citenamefont {Zwierlein}, \citenamefont {Schunck}, \citenamefont {Raupach},\
  and\ \citenamefont {Ketterle}}]{StanNaLi}%
  \BibitemOpen
  \bibfield  {author} {\bibinfo {author} {\bibfnamefont {C.~A.}\ \bibnamefont
  {Stan}}, \bibinfo {author} {\bibfnamefont {M.~W.}\ \bibnamefont {Zwierlein}},
  \bibinfo {author} {\bibfnamefont {C.~H.}\ \bibnamefont {Schunck}}, \bibinfo
  {author} {\bibfnamefont {S.~M.~F.}\ \bibnamefont {Raupach}}, \ and\ \bibinfo
  {author} {\bibfnamefont {W.}~\bibnamefont {Ketterle}},\ }\href {\doibase
  10.1103/PhysRevLett.93.143001} {\bibfield  {journal} {\bibinfo  {journal}
  {Phys. Rev. Lett.}\ }\textbf {\bibinfo {volume} {93}},\ \bibinfo {pages}
  {143001} (\bibinfo {year} {2004})}\BibitemShut {NoStop}%
\bibitem [{\citenamefont {Gacesa}\ \emph {et~al.}(2008)\citenamefont {Gacesa},
  \citenamefont {Pellegrini},\ and\ \citenamefont {C\^ot\'e}}]{GacesaNaLi}%
  \BibitemOpen
  \bibfield  {author} {\bibinfo {author} {\bibfnamefont {M.}~\bibnamefont
  {Gacesa}}, \bibinfo {author} {\bibfnamefont {P.}~\bibnamefont {Pellegrini}},
  \ and\ \bibinfo {author} {\bibfnamefont {R.}~\bibnamefont {C\^ot\'e}},\
  }\href {\doibase 10.1103/PhysRevA.78.010701} {\bibfield  {journal} {\bibinfo
  {journal} {Phys. Rev. A}\ }\textbf {\bibinfo {volume} {78}},\ \bibinfo
  {pages} {010701} (\bibinfo {year} {2008})}\BibitemShut {NoStop}%
\bibitem [{\citenamefont {Schuster}\ \emph {et~al.}(2012)\citenamefont
  {Schuster}, \citenamefont {Scelle}, \citenamefont {Trautmann}, \citenamefont
  {Knoop}, \citenamefont {Oberthaler}, \citenamefont {Haverhals}, \citenamefont
  {Goosen}, \citenamefont {Kokkelmans},\ and\ \citenamefont
  {Tiemann}}]{OberNaLi}%
  \BibitemOpen
  \bibfield  {author} {\bibinfo {author} {\bibfnamefont {T.}~\bibnamefont
  {Schuster}}, \bibinfo {author} {\bibfnamefont {R.}~\bibnamefont {Scelle}},
  \bibinfo {author} {\bibfnamefont {A.}~\bibnamefont {Trautmann}}, \bibinfo
  {author} {\bibfnamefont {S.}~\bibnamefont {Knoop}}, \bibinfo {author}
  {\bibfnamefont {M.~K.}\ \bibnamefont {Oberthaler}}, \bibinfo {author}
  {\bibfnamefont {M.~M.}\ \bibnamefont {Haverhals}}, \bibinfo {author}
  {\bibfnamefont {M.~R.}\ \bibnamefont {Goosen}}, \bibinfo {author}
  {\bibfnamefont {S.~J. J. M.~F.}\ \bibnamefont {Kokkelmans}}, \ and\ \bibinfo
  {author} {\bibfnamefont {E.}~\bibnamefont {Tiemann}},\ }\href {\doibase
  10.1103/PhysRevA.85.042721} {\bibfield  {journal} {\bibinfo  {journal} {Phys.
  Rev. A}\ }\textbf {\bibinfo {volume} {85}},\ \bibinfo {pages} {042721}
  (\bibinfo {year} {2012})}\BibitemShut {NoStop}%
\bibitem [{\citenamefont {Chin}\ \emph {et~al.}(2010)\citenamefont {Chin},
  \citenamefont {Grimm}, \citenamefont {Julienne},\ and\ \citenamefont
  {Tiesinga}}]{ChinFesh}%
  \BibitemOpen
  \bibfield  {author} {\bibinfo {author} {\bibfnamefont {C.}~\bibnamefont
  {Chin}}, \bibinfo {author} {\bibfnamefont {R.}~\bibnamefont {Grimm}},
  \bibinfo {author} {\bibfnamefont {P.}~\bibnamefont {Julienne}}, \ and\
  \bibinfo {author} {\bibfnamefont {E.}~\bibnamefont {Tiesinga}},\ }\href
  {\doibase 10.1103/RevModPhys.82.1225} {\bibfield  {journal} {\bibinfo
  {journal} {Rev. Mod. Phys.}\ }\textbf {\bibinfo {volume} {82}},\ \bibinfo
  {pages} {1225} (\bibinfo {year} {2010})}\BibitemShut {NoStop}%
\bibitem [{\citenamefont {Zirbel}\ \emph
  {et~al.}(2008{\natexlab{a}})\citenamefont {Zirbel}, \citenamefont {Ni},
  \citenamefont {Ospelkaus}, \citenamefont {Nicholson}, \citenamefont {Olsen},
  \citenamefont {Julienne}, \citenamefont {Wieman}, \citenamefont {Ye},\ and\
  \citenamefont {Jin}}]{ZirbelMolODT}%
  \BibitemOpen
  \bibfield  {author} {\bibinfo {author} {\bibfnamefont {J.~J.}\ \bibnamefont
  {Zirbel}}, \bibinfo {author} {\bibfnamefont {K.-K.}\ \bibnamefont {Ni}},
  \bibinfo {author} {\bibfnamefont {S.}~\bibnamefont {Ospelkaus}}, \bibinfo
  {author} {\bibfnamefont {T.~L.}\ \bibnamefont {Nicholson}}, \bibinfo {author}
  {\bibfnamefont {M.~L.}\ \bibnamefont {Olsen}}, \bibinfo {author}
  {\bibfnamefont {P.~S.}\ \bibnamefont {Julienne}}, \bibinfo {author}
  {\bibfnamefont {C.~E.}\ \bibnamefont {Wieman}}, \bibinfo {author}
  {\bibfnamefont {J.}~\bibnamefont {Ye}}, \ and\ \bibinfo {author}
  {\bibfnamefont {D.~S.}\ \bibnamefont {Jin}},\ }\href {\doibase
  10.1103/PhysRevA.78.013416} {\bibfield  {journal} {\bibinfo  {journal} {Phys.
  Rev. A}\ }\textbf {\bibinfo {volume} {78}},\ \bibinfo {pages} {013416}
  (\bibinfo {year} {2008}{\natexlab{a}})}\BibitemShut {NoStop}%
\bibitem [{\citenamefont {Thompson}\ \emph {et~al.}(2005)\citenamefont
  {Thompson}, \citenamefont {Hodby},\ and\ \citenamefont
  {Wieman}}]{ThompOscField}%
  \BibitemOpen
  \bibfield  {author} {\bibinfo {author} {\bibfnamefont {S.~T.}\ \bibnamefont
  {Thompson}}, \bibinfo {author} {\bibfnamefont {E.}~\bibnamefont {Hodby}}, \
  and\ \bibinfo {author} {\bibfnamefont {C.~E.}\ \bibnamefont {Wieman}},\
  }\href {\doibase 10.1103/PhysRevLett.95.190404} {\bibfield  {journal}
  {\bibinfo  {journal} {Phys. Rev. Lett.}\ }\textbf {\bibinfo {volume} {95}},\
  \bibinfo {pages} {190404} (\bibinfo {year} {2005})}\BibitemShut {NoStop}%
\bibitem [{\citenamefont {K\"ohler}\ \emph {et~al.}(2006)\citenamefont
  {K\"ohler}, \citenamefont {G\'oral},\ and\ \citenamefont
  {Julienne}}]{KohlerProdMol}%
  \BibitemOpen
  \bibfield  {author} {\bibinfo {author} {\bibfnamefont {T.}~\bibnamefont
  {K\"ohler}}, \bibinfo {author} {\bibfnamefont {K.}~\bibnamefont {G\'oral}}, \
  and\ \bibinfo {author} {\bibfnamefont {P.~S.}\ \bibnamefont {Julienne}},\
  }\href {\doibase 10.1103/RevModPhys.78.1311} {\bibfield  {journal} {\bibinfo
  {journal} {Rev. Mod. Phys.}\ }\textbf {\bibinfo {volume} {78}},\ \bibinfo
  {pages} {1311} (\bibinfo {year} {2006})}\BibitemShut {NoStop}%
\bibitem [{\citenamefont {Herbig}\ \emph {et~al.}(2003)\citenamefont {Herbig},
  \citenamefont {Kraemer}, \citenamefont {Mark}, \citenamefont {Weber},
  \citenamefont {Chin}, \citenamefont {N\"agerl},\ and\ \citenamefont
  {Grimm}}]{GrimmCs2Science}%
  \BibitemOpen
  \bibfield  {author} {\bibinfo {author} {\bibfnamefont {J.}~\bibnamefont
  {Herbig}}, \bibinfo {author} {\bibfnamefont {T.}~\bibnamefont {Kraemer}},
  \bibinfo {author} {\bibfnamefont {M.}~\bibnamefont {Mark}}, \bibinfo {author}
  {\bibfnamefont {T.}~\bibnamefont {Weber}}, \bibinfo {author} {\bibfnamefont
  {C.}~\bibnamefont {Chin}}, \bibinfo {author} {\bibfnamefont {H.-C.}\
  \bibnamefont {N\"agerl}}, \ and\ \bibinfo {author} {\bibfnamefont
  {R.}~\bibnamefont {Grimm}},\ }\href {\doibase 10.1126/science.1088876}
  {\bibfield  {journal} {\bibinfo  {journal} {Science}\ }\textbf {\bibinfo
  {volume} {301}},\ \bibinfo {pages} {1510} (\bibinfo {year} {2003})},\ \Eprint
  {http://arxiv.org/abs/http://www.sciencemag.org/content/301/5639/1510.full.pdf}
  {http://www.sciencemag.org/content/301/5639/1510.full.pdf} \BibitemShut
  {NoStop}%
\bibitem [{\citenamefont {Spiegelhalder}\ \emph {et~al.}(2010)\citenamefont
  {Spiegelhalder}, \citenamefont {Trenkwalder}, \citenamefont {Naik},
  \citenamefont {Kerner}, \citenamefont {Wille}, \citenamefont {Hendl},
  \citenamefont {Schreck},\ and\ \citenamefont {Grimm}}]{SpeigAllOpt}%
  \BibitemOpen
  \bibfield  {author} {\bibinfo {author} {\bibfnamefont {F.~M.}\ \bibnamefont
  {Spiegelhalder}}, \bibinfo {author} {\bibfnamefont {A.}~\bibnamefont
  {Trenkwalder}}, \bibinfo {author} {\bibfnamefont {D.}~\bibnamefont {Naik}},
  \bibinfo {author} {\bibfnamefont {G.}~\bibnamefont {Kerner}}, \bibinfo
  {author} {\bibfnamefont {E.}~\bibnamefont {Wille}}, \bibinfo {author}
  {\bibfnamefont {G.}~\bibnamefont {Hendl}}, \bibinfo {author} {\bibfnamefont
  {F.}~\bibnamefont {Schreck}}, \ and\ \bibinfo {author} {\bibfnamefont
  {R.}~\bibnamefont {Grimm}},\ }\href {\doibase 10.1103/PhysRevA.81.043637}
  {\bibfield  {journal} {\bibinfo  {journal} {Phys. Rev. A}\ }\textbf {\bibinfo
  {volume} {81}},\ \bibinfo {pages} {043637} (\bibinfo {year}
  {2010})}\BibitemShut {NoStop}%
\bibitem [{\citenamefont {Hadzibabic}\ \emph {et~al.}(2003)\citenamefont
  {Hadzibabic}, \citenamefont {Gupta}, \citenamefont {Stan}, \citenamefont
  {Schunck}, \citenamefont {Zwierlein}, \citenamefont {Dieckmann},\ and\
  \citenamefont {Ketterle}}]{ZoranFifty}%
  \BibitemOpen
  \bibfield  {author} {\bibinfo {author} {\bibfnamefont {Z.}~\bibnamefont
  {Hadzibabic}}, \bibinfo {author} {\bibfnamefont {S.}~\bibnamefont {Gupta}},
  \bibinfo {author} {\bibfnamefont {C.~A.}\ \bibnamefont {Stan}}, \bibinfo
  {author} {\bibfnamefont {C.~H.}\ \bibnamefont {Schunck}}, \bibinfo {author}
  {\bibfnamefont {M.~W.}\ \bibnamefont {Zwierlein}}, \bibinfo {author}
  {\bibfnamefont {K.}~\bibnamefont {Dieckmann}}, \ and\ \bibinfo {author}
  {\bibfnamefont {W.}~\bibnamefont {Ketterle}},\ }\href {\doibase
  10.1103/PhysRevLett.91.160401} {\bibfield  {journal} {\bibinfo  {journal}
  {Phys. Rev. Lett.}\ }\textbf {\bibinfo {volume} {91}},\ \bibinfo {pages}
  {160401} (\bibinfo {year} {2003})}\BibitemShut {NoStop}%
\bibitem [{\citenamefont {Mukaiyama}\ \emph {et~al.}(2004)\citenamefont
  {Mukaiyama}, \citenamefont {Abo-Shaeer}, \citenamefont {Xu}, \citenamefont
  {Chin},\ and\ \citenamefont {Ketterle}}]{MukaiNaDissoc}%
  \BibitemOpen
  \bibfield  {author} {\bibinfo {author} {\bibfnamefont {T.}~\bibnamefont
  {Mukaiyama}}, \bibinfo {author} {\bibfnamefont {J.~R.}\ \bibnamefont
  {Abo-Shaeer}}, \bibinfo {author} {\bibfnamefont {K.}~\bibnamefont {Xu}},
  \bibinfo {author} {\bibfnamefont {J.~K.}\ \bibnamefont {Chin}}, \ and\
  \bibinfo {author} {\bibfnamefont {W.}~\bibnamefont {Ketterle}},\ }\href
  {\doibase 10.1103/PhysRevLett.92.180402} {\bibfield  {journal} {\bibinfo
  {journal} {Phys. Rev. Lett.}\ }\textbf {\bibinfo {volume} {92}},\ \bibinfo
  {pages} {180402} (\bibinfo {year} {2004})}\BibitemShut {NoStop}%
\bibitem [{\citenamefont {Chotia}\ \emph {et~al.}(2012)\citenamefont {Chotia},
  \citenamefont {Neyenhuis}, \citenamefont {Moses}, \citenamefont {Yan},
  \citenamefont {Covey}, \citenamefont {Foss-Feig}, \citenamefont {Rey},
  \citenamefont {Jin},\ and\ \citenamefont {Ye}}]{ChotiaLattice}%
  \BibitemOpen
  \bibfield  {author} {\bibinfo {author} {\bibfnamefont {A.}~\bibnamefont
  {Chotia}}, \bibinfo {author} {\bibfnamefont {B.}~\bibnamefont {Neyenhuis}},
  \bibinfo {author} {\bibfnamefont {S.~A.}\ \bibnamefont {Moses}}, \bibinfo
  {author} {\bibfnamefont {B.}~\bibnamefont {Yan}}, \bibinfo {author}
  {\bibfnamefont {J.~P.}\ \bibnamefont {Covey}}, \bibinfo {author}
  {\bibfnamefont {M.}~\bibnamefont {Foss-Feig}}, \bibinfo {author}
  {\bibfnamefont {A.~M.}\ \bibnamefont {Rey}}, \bibinfo {author} {\bibfnamefont
  {D.~S.}\ \bibnamefont {Jin}}, \ and\ \bibinfo {author} {\bibfnamefont
  {J.}~\bibnamefont {Ye}},\ }\href {\doibase 10.1103/PhysRevLett.108.080405}
  {\bibfield  {journal} {\bibinfo  {journal} {Phys. Rev. Lett.}\ }\textbf
  {\bibinfo {volume} {108}},\ \bibinfo {pages} {080405} (\bibinfo {year}
  {2012})}\BibitemShut {NoStop}%
\bibitem [{\citenamefont {Zirbel}\ \emph
  {et~al.}(2008{\natexlab{b}})\citenamefont {Zirbel}, \citenamefont {Ni},
  \citenamefont {Ospelkaus}, \citenamefont {D'Incao}, \citenamefont {Wieman},
  \citenamefont {Ye},\ and\ \citenamefont {Jin}}]{ZirbelFeshStab}%
  \BibitemOpen
  \bibfield  {author} {\bibinfo {author} {\bibfnamefont {J.~J.}\ \bibnamefont
  {Zirbel}}, \bibinfo {author} {\bibfnamefont {K.-K.}\ \bibnamefont {Ni}},
  \bibinfo {author} {\bibfnamefont {S.}~\bibnamefont {Ospelkaus}}, \bibinfo
  {author} {\bibfnamefont {J.~P.}\ \bibnamefont {D'Incao}}, \bibinfo {author}
  {\bibfnamefont {C.~E.}\ \bibnamefont {Wieman}}, \bibinfo {author}
  {\bibfnamefont {J.}~\bibnamefont {Ye}}, \ and\ \bibinfo {author}
  {\bibfnamefont {D.~S.}\ \bibnamefont {Jin}},\ }\href {\doibase
  10.1103/PhysRevLett.100.143201} {\bibfield  {journal} {\bibinfo  {journal}
  {Phys. Rev. Lett.}\ }\textbf {\bibinfo {volume} {100}},\ \bibinfo {pages}
  {143201} (\bibinfo {year} {2008}{\natexlab{b}})}\BibitemShut {NoStop}%
\bibitem [{\citenamefont {Hodby}\ \emph {et~al.}(2005)\citenamefont {Hodby},
  \citenamefont {Thompson}, \citenamefont {Regal}, \citenamefont {Greiner},
  \citenamefont {Wilson}, \citenamefont {Jin}, \citenamefont {Cornell},\ and\
  \citenamefont {Wieman}}]{HodbyPhaseSp}%
  \BibitemOpen
  \bibfield  {author} {\bibinfo {author} {\bibfnamefont {E.}~\bibnamefont
  {Hodby}}, \bibinfo {author} {\bibfnamefont {S.~T.}\ \bibnamefont {Thompson}},
  \bibinfo {author} {\bibfnamefont {C.~A.}\ \bibnamefont {Regal}}, \bibinfo
  {author} {\bibfnamefont {M.}~\bibnamefont {Greiner}}, \bibinfo {author}
  {\bibfnamefont {A.~C.}\ \bibnamefont {Wilson}}, \bibinfo {author}
  {\bibfnamefont {D.~S.}\ \bibnamefont {Jin}}, \bibinfo {author} {\bibfnamefont
  {E.~A.}\ \bibnamefont {Cornell}}, \ and\ \bibinfo {author} {\bibfnamefont
  {C.~E.}\ \bibnamefont {Wieman}},\ }\href {\doibase
  10.1103/PhysRevLett.94.120402} {\bibfield  {journal} {\bibinfo  {journal}
  {Phys. Rev. Lett.}\ }\textbf {\bibinfo {volume} {94}},\ \bibinfo {pages}
  {120402} (\bibinfo {year} {2005})}\BibitemShut {NoStop}%
\bibitem [{\citenamefont {Williams}\ \emph {et~al.}(2006)\citenamefont
  {Williams}, \citenamefont {Nygaard},\ and\ \citenamefont
  {Clark}}]{WillMolTheory}%
  \BibitemOpen
  \bibfield  {author} {\bibinfo {author} {\bibfnamefont {J.~E.}\ \bibnamefont
  {Williams}}, \bibinfo {author} {\bibfnamefont {N.}~\bibnamefont {Nygaard}}, \
  and\ \bibinfo {author} {\bibfnamefont {C.~W.}\ \bibnamefont {Clark}},\ }\href
  {\doibase 10.1088/1367-2630/8/8/150} {\bibfield  {journal} {\bibinfo
  {journal} {New J. Phys.}\ }\textbf {\bibinfo {volume} {8}},\ \bibinfo {pages}
  {150} (\bibinfo {year} {2006})}\BibitemShut {NoStop}%
\bibitem [{\citenamefont {Deuretzbacher}\ \emph {et~al.}(2008)\citenamefont
  {Deuretzbacher}, \citenamefont {Plassmeier}, \citenamefont {Pfannkuche},
  \citenamefont {Werner}, \citenamefont {Ospelkaus}, \citenamefont {Ospelkaus},
  \citenamefont {Sengstock},\ and\ \citenamefont {Bongs}}]{DeuretzMolOptLatt}%
  \BibitemOpen
  \bibfield  {author} {\bibinfo {author} {\bibfnamefont {F.}~\bibnamefont
  {Deuretzbacher}}, \bibinfo {author} {\bibfnamefont {K.}~\bibnamefont
  {Plassmeier}}, \bibinfo {author} {\bibfnamefont {D.}~\bibnamefont
  {Pfannkuche}}, \bibinfo {author} {\bibfnamefont {F.}~\bibnamefont {Werner}},
  \bibinfo {author} {\bibfnamefont {C.}~\bibnamefont {Ospelkaus}}, \bibinfo
  {author} {\bibfnamefont {S.}~\bibnamefont {Ospelkaus}}, \bibinfo {author}
  {\bibfnamefont {K.}~\bibnamefont {Sengstock}}, \ and\ \bibinfo {author}
  {\bibfnamefont {K.}~\bibnamefont {Bongs}},\ }\href {\doibase
  10.1103/PhysRevA.77.032726} {\bibfield  {journal} {\bibinfo  {journal} {Phys.
  Rev. A}\ }\textbf {\bibinfo {volume} {77}},\ \bibinfo {pages} {032726}
  (\bibinfo {year} {2008})}\BibitemShut {NoStop}%
\end{thebibliography}

%

\end{document}